\documentclass[12pt]{article}

\usepackage{amsmath}
\usepackage{amssymb}
\usepackage{graphicx}
\usepackage{subfigure}

\def\middlespace {\smallskipamount=5.625pt plus1.5pt minus1.5pt
                  \medskipamount=11.25pt plus3pt minus3pt
                  \bigskipamount=22.5pt plus6pt minus6pt
                  \normalbaselineskip=22.5pt plus0pt minus0pt
                  \normallineskip=1pt
                  \normallineskiplimit=0pt
                  \jot=5.625pt
                  {\def\smallskip {\vskip\smallskipamount}}
                  {\def\medskip   {\vskip\medskipamount}}
                  {\def\bigskip   {\vskip\bigskipamount}}
                  {\setbox\strutbox=\hbox{\vrule
                    height15.75pt depth6.75pt width 0pt}}
                  \parskip 11.25pt
                  \normalbaselines}

\begin{document}


\begin{center}
 { \Large {\bf Nonlinear Quantum Mechanics, the Superposition }}

\smallskip

{\Large {\bf  Principle, and the Quantum Measurement Problem}}

\vskip 0.2 in

\smallskip

\bigskip

\bigskip

\bigskip

{\large{\bf Kinjalk Lochan\footnote{e-mail address: kinjalk@tifr.res.in} and T. P. Singh\footnote{e-mail address: tpsingh@tifr.res.in} }}

\medskip

{\it Tata Institute of Fundamental Research,}\\
{\it Homi Bhabha Road, Mumbai 400 005, India.}\\
\medskip

\vskip 0.5cm
\end{center}

\vskip 1.0 in


\begin{abstract}
\noindent 
There are four reasons why our present knowledge and understanding of quantum mechanics could be regarded as incomplete. Firstly, the principle of linear superposition has not been experimentally tested for position eigenstates of  objects having more than about a thousand atoms. Secondly, there is no universally agreed upon explanation for the process of quantum measurement. Thirdly, there is no universally agreed upon explanation for the observed fact that macroscopic objects are not found in superposition of position eigenstates. Fourthly, and perhaps most importantly, the concept of time is classical and hence external to quantum mechanics : there should exist an equivalent reformulation of the theory which does not refer to an external classical time. In this paper we argue that such a reformulation is the limiting case of a nonlinear quantum theory, with the nonlinearity becoming important at the Planck mass scale. Such a nonlinearity can provide insights into the problems mentioned above. We use a physically motivated model for a nonlinear Schr\"{o}dinger equation to show that nonlinearity can help in understanding quantum measurement. We also show that while the principle of linear superposition holds to a very high accuracy for atomic systems, the lifetime of a quantum superposition becomes progressively smaller, as one goes from microscopic to macroscopic objects. This can explain the observed absence of position superpositions in macroscopic objects [lifetime is too small]. It also suggests that ongoing laboratory experiments maybe able to detect the finite superposition lifetime for mesoscopic objects, in the foreseeable future.     

\noindent 

\vskip 1.0 in

\end{abstract}

\newpage

\section{Introduction}

The principle of linear superposition is a central tenet of quantum mechanics, and is not contradicted by any experiment to date. It has been tested successfully in the laboratory for molecules as large as $C_{60}$ (fullerene), consisting of sixty atoms \cite{zeillinger0}, \cite{zeillinger}. Does the principle hold for even larger molecules? Is there a limit to how large an object can be, before the principle breaks down, or does it hold for objects containning arbitrary number of atoms? We do not know  the experimental answer to these questions.  

If the principle does hold for objects of arbitrarily large size, as would be the case in standard quantum theory, then how does one understand the apparent breakdown of superposition during a quantum measurement? One possible answer, within linear quantum theory, is that indeed the breakdown of superposition during a measurement is only an apparent phenomenon, not a real one. When a quantum system interacts with the apparatus, the two together form a macroscopic system. In a macroscopic system, because of its interaction with the environment, the superposition decoheres on a very short time scale, that is, the interference between different basis states is almost completely destroyed. The observer detects only one amongst the many outcomes because the other outcomes have been realized in other Universes - this is Everett's many worlds interpretation, crucial to understanding the physics of quantum measurement, if linear superposition is assumed to hold for large objects of all sizes, and hence also during a quantum measurement.      

Decoherence and many worlds is also crucial to understanding why a macroscopic object is never observed in more than one  place at the same time. The explanation is the same as that for the result of a quantum measurement. If an initial state could be prepared in which the object is in more than one position at the same time, the superposition will decohere extremely rapidly. The object will be in different positions in different branches of the Universe, but in our branch only one position will be detected.

Thus in standard quantum theory, linear superposition holds on all scales, and decoherence and many worlds is one possible way of understanding quantum measurement and the absence of macroscopic superpositions. While decoherence is a completely physical process which has also been experimentally verified in the laboratory, the same unfortunately cannot be said about the branching in the many-worlds scenario. How do we experimentally verify that these other branches do indeed exist? As of today, we do not know the answer. Indeed, by the very nature of its construction, the various branches of the many worlds Universe are not supposed to interact with each other. The scenario should thus be treated as a hypothesis, until it can be verified.    

Of course, within linear quantum theory, many-worlds is not the only possible explanation for quantum measurement, a prominent contender being Bohmian mechanics. This is one example of what are often called hidden variable theories, wherein the wave-function evolving according to the 
Schr\"{o}dinger equation provides only partial information about a quantum system. The description is completed by specifying the actual positions of particles which evolve according to a guiding equation where the velocities of the particles are expressed in terms of the wave-function. The evolution of the system is deterministic, including during quantum measurement, while the wave-function evolves linearly. It is not clear how Bohmian mechanics could be experimentally distinguished from the standard quantum theory.  

Let us leave the discussion of these issues aside for a moment, and address a completely different incompleteness in our understanding of quantum mechanics. This is the presence of an external classical time in the theory. Classical time is part of a classical spacetime geometry, which is produced by classical matter fields, which in turn are a limiting case of quantum fields. A fundamental formulation of the theory should not have to depend on its own limiting case. Hence such a time cannot be a part of a fundamental formulation of the theory - there must exist an equivalent reformulation of the theory which does not make reference to classical time. We will argue in the next section that such a reformulation is the limiting case of a nonlinear quantum theory, with the nonlinearity becoming significant only in the vicinity of the Planck mass scale. Away from the Planck mass scale, for objects with much smaller masses, the theory is linear to a very high approximation. For objects with masses much larger than Planck mass the theory goes over to classical mechanics. 

The conclusion mentioned in the previous paragraph, namely that quantum theory is inherently nonlinear, is a consequence of the relation of the theory to the structure of spacetime. But it becomes immediately apparent that this nonlinearity has a bearing on the questions raised earlier in the Introduction. Thus, the principle of linear superposition is not an exact feature of quantum theory, but only an approximate one. For microscopic systems made of a few atoms, the linear approximation holds to an extremely high accuracy. The lifetime of a quantum superposition is astronomically large, so that linear superposition appears to be an exact property for the quantum mechanics of microscopic systems. As the number of atoms in an object increases, the lifetime of a quantum superposition will start to decrease. When the number of atoms $N$ in the object becomes large [by large we will mean $N$ greater than about $10^{18}$, which is the number of atoms in a Planck mass] the object becomes macroscopic, and the superposition lifetime becomes unmeasurably small (by the standards of today's technology). We propose a new domain, the mesoscopic domain, in which the number of atoms in an object is neither microscopic, nor macroscopic. By microscopic we will approximately mean $N$ less than about $10^{3}$. As one goes from the microscopic domain, towards the mesoscopic domain, the superposition lifetime will smoothly decrease, and one naturally expects that there will be a range of values of $N$ for which the superposition lifetime will neither be astronomically large nor unmeasurably small. These values of $N$ will clearly be of interest from an experimental point of view, so that these ideas can be put to test. While we leave a precise estimation of the range of such $N$ for future work, we will derive some illustrative values later in the paper. Possible connection with ongoing experiments will be discussed in Section 6.

The fact that the theory becomes nonlinear on the Planck mass scale could also have a bearing on the quantum measurement problem. When a quantum system interacts with a measuring apparatus, their further joint evolution is governed by a nonlinear Schr\"{o}dinger equation. The nonlinearity can cause a breakdown of the initial superposition, and drive the system to one specific outcome. Which particular outcome will be realized depends on the value taken by an [effectively] random parameter. If the random variable has an appropriate probability distribution, the outcomes obey the Born probability rule. The principle task of this paper is to use a model nonlinear equation, and demonstrate how a quantum measurement can be explained.   The same reasoning then also explains the absence of superpositions for macroscopic objects.

A key feature of this analysis is that the nonlinear equation yields a quantitative  estimate for the lifetime of a quantum superposition. This lifetime should be compared with the time-scale over which decoherence takes place in a system, in order to decide which of the two [decoherence or nonlinearity] are more important in a given situation. Our thesis is that the measurement process can be explained by the nonlinearity, and we need not invoke many worlds. This thesis can be put to experimental test in the laboratory, unlike many worlds. What is most important for us is that we do not introduce the nonlinearity in an ad hoc manner to explain quantum measurement. The origin of the nonlinearity lies in the relation between quantum mechanics and spacetime structure - its impact on quantum measurement is a natural by-product. The nonlinear equation we will use in Section 5 is a model, in the sense that it is not rigorously derived from an underlying mathematical description of the required reformulation of quantum mechanics. Nonetheless, it is motivated by considerations of the structure that such a reformulation could be expected to take.

Contrary to a general impression, non-linear Schr\"{o}dinger equations need not always be ad hoc. An excellent example of this is the Doebner-Goldin equation, which arises very naturally during the construction of representations of current algebras. Interestingly, and for reasons not understood at present, our nonlinear equation is very similar to the D-G equation. In Section 4 we recall the simple and elegant derivation of the D-G equation - to some degree this serves also as a motivation for the particular nonlinear equation that we consider. 

The idea that the breakdown of superposition during a quantum measurement comes about because of a dynamical modification of the Sch\"{o}dinger equation is not new. It can perhaps be traced back to the early work of Bohm and Bub \cite{bohmbub} developed in the context of hidden variable theories. It was discussed again by Pearle \cite{pearle}. What is new in this paper is the proposal that the nonlinearity has a genuine origin in quantum theory itself, when one considers its relation to spacetime structure. [It is worth mentionning here though that a similar suggestion was made by Feynman \cite{feynman}.] The key aspects of nonlinearity induced measurement were explained in a nice and simple model by Grigorenko \cite{grigorenko}, which we recall in Section 3. Our analysis in Section 5, using our nonlinear Schr\"{o}dinger equation, is similar to Grigorenko's analysis. Also in Section 5, we discuss the often debated issue of the relation between nonlinearity and superluminality.

There are various other models of dynamically induced collapse, for example the work of Adler \cite{adler}, Ghirardi, Rimini and Weber \cite{grw}, Diosi \cite{diosi}, and Penrose \cite{penrose}. These and other models of dynamical collapse have been briefly reviewed in \cite{singh2007}. A general discussion of nonlearity in quantum mechanics is in Weinberg \cite{weinberg}.

In  the next Section we argue why quantum theory becomes nonlinear on the Planck mass scale.

\section{Spacetime structure and nonlinear quantum mechanics}

The concept of time evolution is of course central to any dynamical theory, and in particular to quantum mechanics. In standard quantum mechanics time, and spacetime, are taken as given. But the presence of time in the theory is an indicator of a fundamental incompleteness in our understanding, as we now elaborate. Time cannot be defined without an external gravitational field [this could be flat Minkowski spacetime, or a curved spacetime]. The gravitational field is of course classical. Thus the picture is that an external spacetime manifold and an overlying gravitational field must be given, before one can define time evolution in quantum theory. 

This classical gravitational field is produced by classical matter, in accordance with the laws of classical general relativity. If the Universe did not have any classical matter, there would be no classical spacetime metric. There is an argument due to Einstein, known as the Einstein hole argument \cite{Einstein}, \cite{christian}, according to which, if the spacetime manifold is to have a well-defined point structure, there must reside on the manifold a physically determined (classical) spacetime metric. If there are no classical matter fields, the spacetime metric will undergo quantum fluctuations - these destroy the underlying manifold. Thus, in the absence of background classical matter, one cannot talk of the usual time evolution in quantum theory. Nonetheless, there ought to exist a formulation of quantum mechanics which describes even such circumstances. Because one can well imagine such a situation [say immediately after the Big Bang] in which there are no classical matter fields at all. How is one to describe the `quantum dynamics' of such quantum mater fields? Such a (re)formulation of the theory must be exactly equivalent to the original theory, in the sense that standard quantum theory must follow from it, as and when external classical matter fields, and hence a background classical spacetime geometry, exist.         

We now come to the central thesis of this paper. We will argue that this reformulation is the limiting case of a nonlinear quantum theory, with the nonlinearity becoming important at the Planck mass scale \cite{singhncomm}. Of course this nonlinear theory also does not refer to any external classical time. Standard quantum mechanics then becomes the limiting case of a nonlinear quantum mechanics, both of which refer to an external time, and the nonlinearity is significant only in the vicinity of the Planck mass scale 
\begin{equation}
m_{Pl}\sim (\hbar c/G)^{1/2}\sim 10^{-5} {\rm gms} \sim 10^{18} {\rm atoms}. 
\end{equation}
To establish this argument consider a collection of quantum mechanical  particles, whose dynamics is being described with respect to an external time. Next, let us imagine that this external background time is no longer available, and furthermore, let us assume that the total mass-energy of this collection of particles is much less than Planck mass : $m_{total}\ll m_{Pl}$. We take this to be the approximation $m_{Pl}\rightarrow\infty$, and since $m_{Pl}\propto G^{-1/2}$, this is also the approximation $G\rightarrow 0$. Physically, and plausibly, this means that if the total mass energy of the quantum mechanical particles is much less than Planck mass, their gravitational effect is ignorable. The dynamics corresponds to the situation where, had an external flat spacetime been available, motion of the particles takes place without gravitationally distorting the background.  

Consider next the situation that the total mass-energy of the particles becomes comparable to Planck mass. Their gravitational field becomes important now. The quantum gravitational effect of the particles on their own dynamics can no longer be ignored. The effect feeds back on itself iteratively, and the dynamics is evidently nonlinear, in the sense that the evolution of the state depends on the state itself. [This is analogous to the situation in classical general relativity, where gravity acts as a source for itself, making general relativity a nonlinear theory.] Had an external spacetime been available, this dynamics would correspond to one where self-gravity effects distort the background and the evolution equation is nonlinear. In the nonrelativistic limit, this would be a nonlinear Schr\"{o}dinger equation. There is no analogue of this in standard quantum theory. It is also evident from these considerations that the quantum gravitational effect feeds back on itself, and hence quantum gravity is a nonlinear theory on the Planck energy scale. Again, this is in contrast to a conventional approach to quantum gravity such as quantum general relativity described by the Wheeler deWitt equation, which is a linear theory.   

We should carefully examine as to what the connection between the reformulation and the nonlinearity is. For instance, can one not see, in standard quantum theory itself, by an argument identical to the one above, that the effect of self-gravity will make the Schr\"{o}dinger equation nonlinear on the Planck mass scale? Indeed that would be the case, but we simply do not know how to formulate that in a self-consistent, non-perturbative way, starting from the linear quantum theory. The usual approach to quantization is to apply the linear rules of quantization to a classical theory. If the quantum gravitational feedback becomes important, how does one generalize the linear Wheeler deWitt equation  to a nonlinear equation? Correspondingly, how does one take into account the nonlinear feedback in the Schr\"{o}dinger equation? There are no prescriptions to address these questions. On the other hand, when we start from the reformulation, we are no longer restricted by the language and formalism of standard linear quantum theory. In all likelihood, a new formalism will have to be implemented (the possible use of noncommutative geometry has been suggested in \cite{singhncomm}). This formalism will probably by itself pave the way for a natural generalization, as happens for instance in the transition from special to general relativity [Lorentz invariance generalised to general covariance].   
 
A crucial fall-out of the approach to nonlinearity from the issue of reformulation of quantum mechanics is the following. It has been said that even if self-gravity effects were to be taken into consideration in quantum theory, and some way of introducing nonlinearity on Planck mass scale were to be found, the effects will be utterly negligible in laboratory physics. This is absolutely correct for the following reason. A Planck mass size object in the laboratory is much, much larger in size compared to its Schwarzschild radius [the former would typically be of the order of a micron, and the latter of the order of a Planck length]. Self-gravity, if it has to significantly distort the background and effect the object's motion, would be relevant only if the object's size is comparable to its Schwarzschild radius. 

However, when we start looking for a reformulation of quantum mechanics, and its nonlinear generalization, there is nothing to prevent the suggestion that the gravitational field has additional components [for instance an antisymmetric part] which are important on the micron length scale for a Planck mass object [this is not ruled out by experiment]. By important we mean that these components could make the effects of self-gravity relevant for laboratory physics, in a manner not anticipated thus far. It is nonlinear effects of this nature that we will study in this paper, with the help of a model, in Section 5. 

There is a compelling reason to believe that in the reformulation, and in its nonlinear generalisation, spacetime and gravity will have additional structure. It is simply the original reason for the necessity of the reformulation - the loss of the point structure of the Riemannean spacetime manifold, and the attached symmetric spacetime metric that comes with it. Since these cannot be part of the reformulation, and yet must be recovered from the reformulation in a certain approximation [classical limit], whatever mathematical structure describes the reformulation must be more general than a spacetime manifold and the accompanying symmetric metric. The same holds for the nonlinear generalisation of the reformulation, which becomes important for $m\sim m_{Pl}$, a domain different from the strictly classical limit $m\gg m_{Pl}$. The classical limit in our picture is the limit $m_{Pl}\rightarrow 0$, which corresponds to $\hbar\rightarrow 0, G\rightarrow\infty$.  

Having argued that the nonlinearity is important, we would now like to study the implications of a nonlinear quantum mechanics for the measurement problem. Before we do so, we illustrate the key features with a toy model.

\section{A nonlinear Schr\"{o}dinger equation and quantum measurement}

In this section we examine one class of modified Schr\"{o}dinger equations studied by Grigorenko  \cite {grigorenko}. No fundamental physical 
reason is given for choosing such a class of equations except that it serves the purpose of modelling collapse of the wave-function during a measurement.

Any equation modelling a non-destructive measurement process must be norm-preserving. It must be nonlinear in the wave-function of the system 
so that it facilitates the collapse of the wave-function from a state of superposition to an eigenstate of the measured quantity. The choice of
 eigenstate to collapse into must be random but in accordance with the Born probability rule, as has been experimentally observed. Grigorenko's 
equation satisfies these requirements when an operator $U$, modelling the action of the measuring apparatus, is chosen to be in the following class, first discovered by Gisin \cite{Gisin}
\begin{equation}
i \hbar \frac{\partial |\psi>}{\partial t} = H |\psi> + (1 - P_{\psi}) U |\psi>
\label{grigoeqn}.
\end{equation}

Here \(P_\psi \) is the projection operator, $H$ is the standard Hermitian Hamiltonian, $U$ is any arbitrary linear or nonlinear operator, not 
necessarily Hermitian if linear. It is seen that (\ref{grigoeqn}) is norm-preserving, when $ d<\psi|\psi>/dt $ is calculated and found to be zero. 
Addition of a term A(1 - \(P_\psi \)) to the Hamiltonian makes no difference to (\ref{grigoeqn}) where $A$ is an arbitrary operator since action of this
 term on normalized wave-function gives zero. Thus different Hamiltonians result in the same equations of motion for the particle.

For some Hamiltonians (\ref{grigoeqn}) is able to facilitate collapse of a superposition to a single eigenstate.
In the simplest case, the Hermitian part of the Hamiltonian is taken to be zero and a linear $U$ is chosen such that \( U = - U^\dagger \). No generality is 
lost in using such a form for $U$. The operator may be written as the sum of a Hermitian operator R and an anti-Hermitian operator S. Modifying the
 wave-function as \(|\psi'>\ = e^{-i\kappa t} |\psi> \) and the operators as \(H' = e^{-i\kappa t} (H + R) e^{i\kappa t} \) and 
\(U' = e^{-i\kappa t} S e^{i\kappa t} \) will let the modified operators satisfy (\ref{grigoeqn}) if \(\kappa = \int_{0}^{t} <\psi|R|\psi>dt\).

To see how collapse occurs, the following form for $U$ may be taken: \(U = \sum_n i\gamma_n |\phi_n><\phi_n| \) where $\phi_n$ are the eigenstates 
of an operator A which corresponds to the observable being measured. The \(q_n \)s are random real variables associated with each $|\phi_n>$ with some
 probability distribution. They do not change with time after the onset of measurement. $\gamma$ is a real coupling constant that could be dependent
 on the number of degrees of freedom of the entire system, so that it is 'turned on' at the start of measurement when the apparatus begins to interact
 with the quantum system. Taking \(|\psi> = \sum_n a_n |\phi_n>\) and the said form of $U$ in (\ref{grigoeqn}), the following relations are obtained,

\begin{equation}
\frac{da_n}{dt} = \gamma a_n (q_n - \sum_n q_n |a_n|^2) ,
\end{equation}

\begin{equation}
\frac{d}{dt} \ln \left( \frac{|a_i|^2}{|a_j|^2} \right) = 2 \gamma (q_i - q_j) .
\label{ratioeqn}
\end{equation}

When $U$ is not time-dependent, (\ref{grigoeqn}) has the following exact solution,

\begin{equation}
|\psi(t)> = \frac{e^{-i\:(H_0 + U) t} |\psi(t_0)>}{\sqrt{<\psi(t_0)|e^{-2iUt}|\psi(t_0)>}} ,
\end{equation}

\begin{equation}
|a_i(t)| = \frac{|\exp (\gamma q_i t) a_i(t_0)|}{\sqrt{(\sum_n \exp(2 \gamma q_n t) |a_n(t_0)|^2)}} .
\label{aieqn}
\end{equation}
Therefore,
\begin{equation}
\frac{|a_i(t)|}{|a_j(t)|} = \exp (\gamma (q_i-q_j) t)\frac{| a_i(t_0)|}{|a_j(t_0)|}.
\label{ratio1}
\end{equation}

From (\ref{ratioeqn}) and (\ref{aieqn}), it is seen that the state with the largest value of $q$ has its amplitude growing the fastest. 
It is also noted that an eigenstate with zero probability amplitude at the start of the measurement continues with it irrespective of 
the random variable associated with it. Since Grigorenko's equation is a norm preserving one, only the state with the largest 
value of q survives after the measurement.

 Let $ q_i $ be the largest among all $ q_n $ ; then the ratio (\ref{ratio1}) would grow in favor of $ |a_i(t)| $, on the time scale,  
\begin{equation}
 \tau = \frac{1}{\gamma (q_i-q_j)}.
\end{equation}
We can see that the time scale of collapse is inversely proportional to the coupling constant $\gamma$. Larger the coupling strength, shorter would 
be the collapse time.

The Grigorenko equation also allows a derivation of the Born probability rule in the following manner. If $p_i$ is the probability of the 
wave-function collapsing into state \(|\phi_i> \), it is the same as the probability of \(q_i\) being greater than all other \(q_n\), as 
the state with the largest value of q is seen to survive,

\begin{equation}
p_i = \int .. \int \omega(q_i) dq_i \prod_{n \neq 1} \theta(q_i - q_n) \omega(q_n) dq_n
.\label{p1}
\end{equation}

The Born probability rule requires that \(p_i = |<\psi(t_0)|\phi_i>|^2 \). The following probability distribution when used in (\ref{p1}) 
is consistent with the Born probability rule :

\begin{equation}
\omega (q_n) = |<\psi (t_0)|\phi_n>|^2 e^{|<\psi (t_0)|\phi_n>|^2 q_n},
\label{ref2eqn}
\end{equation}
for random variables $q_n$, distributed along $(-\infty,0 ].$

However, this probability distribution is not uniquely determined. Any change in variables in the integral (6), which does not change the value of the 
integral and does not change the projection property of (1), provides other distributions with correct outcome probabilities.

\subsection{Phase as a possible choice of random variable}
Random variables associated with each eigenstate play an important role in deriving the Born probability rule. A natural choice for them are 
the respective phases of the eigenstates. Phases $\chi_n$ are uniformly distributed in [0, 2$\pi$] and \(u_n = \ln (\chi_n/ 2 \pi)\) are random 
variables with probability distribution as follows:

\begin{equation}
\omega (u_n) = e^{u_n}.
\label{ref3eqn}
\end{equation}
Ignoring the Hermitian part of Hamiltonian we see that the phases remain constant through measurement.

The random variables \(q_n\) are fixed at the time of onset of measurement and their probability distributions depend on the initial state
 of the system to yield the Born rule. Transforming \(q_n\) as \(q_n = u_n/|<\psi(t_0)|\phi_n>|^2\) removes the said dependence and probability
 distribution of \(u_n\) is given as \(\omega (u_n) = e^ {u_n}\). Using these the following equation is obtained,

\begin{equation}
\frac{d|a_i|^2}{dt}  = 2 \gamma \left(u_i - |a_i|^2 \sum_i u_i\right).
\end{equation}
Using the transformation $ (q_n \Rightarrow -1/q_n)$, one avoids a singularity when \(|a_n|^2 = 0\). \(u_n\) formed from the phases has an 
exponential distribution as required above, thus confirming that phases can be used to form the random variables required in Grigorenko's model.

Thus the above scheme of Grigorenko satisfactorily models measurement. But it is useful to note that this is not the only one that does. Also, 
it is desirable to find rigorous reasoning that dictates the form of the modification to the Schr\"{o}dinger equation. We will now recall the Doebner-Goldin nonlinear equation, which has an elegant and compelling theoretical origin, and which closely resembles the equation we will use for modelling measurement, in Section 5.

\section{The nonlinear Schr\"{o}dinger equation of Doebner and Goldin}
Doebner and Goldin \cite{DG} were led to their nonlinear equation by considering representations 
of the current algebra formulation of non-relativistic quantum mechanics. To understand this formulation let us consider the time-evolution of 
the wave-function of a single particle given by the Schr\"{o}dinger equation.
\begin{equation}
i \hbar \frac{\partial \psi}{\partial t} = - \frac{\hbar^2}{2m} \nabla^2 \psi + V \psi
\label{seqn} .
\end{equation}

Defining the quantities probability density $\rho$ and current density {\bf j} of the wave-function as 
\begin{equation}
\rho = \psi^* \psi
\label{rho} ,
\end{equation}

\begin{equation}
{\bf j} = \frac{\hbar}{2mi} ( \psi^* \nabla \psi - \psi \nabla \psi^*)
\label{jeqn} ,
\end{equation}
we see that (\ref{seqn}) can  be written in terms of densities as the continuity equation,
\begin{equation}
\frac{\partial \rho}{\partial t} = - {\bf \nabla. j}
\label{ceqn} .
\end{equation}

[One may also arrive at the Schr\"{o}dinger equation starting from the continuity equation if the above definitions of probability current 
and density are used.
Expanding (\ref{ceqn}) using (\ref{rho}) and (\ref{jeqn}) one obtains the following:

\begin{eqnarray}
 \psi^* \frac{\partial \psi}{\partial t} + \psi \frac{\partial  \psi^*}{\partial t} &=& -\frac{\hbar}{2im} \nabla . ( \psi^* \nabla \psi - \psi \nabla \psi^*),\\
2\:Re ( \psi^* \frac{\partial \psi}{\partial t}) &=& \frac{\hbar}{2im}(-\psi^* \nabla^2 \psi + \psi \nabla^2  \psi^*) = \frac{\hbar}{m} \:Im ( \psi^* \nabla^2 \psi) 
\label{cceqn},
\end{eqnarray}
\begin{equation}
 \:Re ( \psi^* \frac{\partial \psi}{\partial t})=\frac{\hbar}{2m}\:Im ( \psi \nabla^2 \psi^*). 
\end{equation}
We can show 
\begin{equation}
Re(ab) = \frac{\hbar}{2m}\:Im(ac) \Rightarrow b = -i\frac{\hbar}{m}\:c,
\end{equation}
leading to
\begin{equation}
\frac{\partial \psi^*}{\partial t} = -i\frac{\hbar}{2m}\:\nabla^2 \psi^*,
\end{equation}
and hence
\begin{equation}
i\hbar \frac{\partial \psi}{\partial t} = -\frac{\hbar^2}{2m}\:\nabla^2 \psi.
\end{equation}
Adding to and subtracting from (\ref{cceqn}) a term of the form V$\psi$ where V is a real function will make no difference to the continuity 
equation and hence will lead to the derivation of the more general form (\ref{seqn}).]

Doebner and Goldin have studied the non-relativistic wave-function as a field writing the field theory in terms of current algebras to see 
if this yields any new insights or interesting results.   
The wave-function is second quantized and using the commutation relations 
\begin{equation}
[ \psi^*({\bf x}, t) , \psi({\bf y}, t)]_\pm = \delta({\bf x} - {\bf y}),
\end{equation}
we obtain the following commutation relations between $\rho$ and {\bf j},
\begin{equation}
[\rho({\bf x}, t) , \rho({\bf y}, t)] = 0
\label{com1},
\end{equation}

\begin{equation}
[\rho({\bf x}, t) , {\bf j}_k({\bf y}, t)] = - i \frac{\partial}{\partial x^k}[ \delta({\bf x} - {\bf y}) \rho ({\bf x}, t)]
\label{com2},
\end{equation}

\begin{equation}
[{\bf j}_i({\bf x}, t) , {\bf j}_k({\bf y}, t)] = - i \frac{\partial}{\partial x^k}[ \delta({\bf x} - {\bf y}) {\bf j}_i ({\bf x}, t)] +
 i \frac{\partial}{\partial y^i}[ \delta({\bf x} - {\bf y}) {\bf j}_k ({\bf y}, t)].
\label{com3}
\end{equation}
$\rho$ and {\bf j} are averaged over space using Schwartz functions $f$ and components of {\bf g}, a vector field over $R^3$ and the algebra of operators is obtained,

\begin{equation}
\rho_{op} (f) = \int{\rho ({\bf x}, t) f ({\bf x}) d{\bf x}},
\end{equation}

\begin{equation}
j_{op} ({\bf g}) = \int{{\bf j} ({\bf x} , t) {\bf.} {\bf g}({\bf x}) d{\bf x}},
\end{equation}

\begin{equation}
[\rho_{op} (f_1), \rho_{op} (f_2)] = 0
\label{c1},
\end{equation}

\begin{equation}
[\rho_{op} (f), j_{op} ({\bf g})] = i \frac{\hbar }{m} \rho_{op} ({\bf g.\nabla} f)
\label{c2},
\end{equation}

\begin{equation}
[j_{op} ({\bf g_1}), j_{op} ({\bf g_2})] = - i \frac{\hbar}{m} j_{op} ({\bf g_1 \nabla.g_2 - g_2 \nabla.g_1})
\label{c3}.
\end{equation}
This algebra represents a group which is the semi-direct product of the group of diffeomorphisms of space and the space of Schwartz functions.
 The group of diffeomorphisms is the most general symmetry group of the configuration space of a particle \cite{goldinref}.

Now we ask if the time-evolution of $\rho$ and {\bf j} can completely describe a quantum mechanical system just the way a particle's position 
and momentum and their time evolution can describe its dynamics completely. In other words, it should be possible to express any physical 
observable associated with the system in terms of these two quantities. The answer is shown to be yes in \cite{Sharp}, that $\rho$ and {\bf j} form 
a complete irreducible set of coordinates with which to describe the system.

Doebner and Goldin have taken the commutation relations (\ref{c1}), (\ref{c2}), (\ref{c3}) between $\rho$ and {\bf j} as input from quantum
 mechanics and asked what dynamical equation $\rho$ and {\bf j} must obey so that their commutation relations are preserved. It must be noted that
 neither the definitions of the quantities $\rho$ and {\bf j}, nor their commutation relations, depend on the original Schr\"{o}dinger equation.
 Hence it is logically consistent to start from commutation relations and arrive at an equation for the dynamics of a particle that is different
 from the Schr\"{o}dinger equation.

A possible one-particle representation of the algebra in (\ref{c1}), (\ref{c2}), (\ref{c3}) is given by the action of the following operators on 
the wave-function of the particle at a particular time,
\begin{equation}
\rho_{op} (f) \psi ({\bf x}) = f ({\bf x}) \psi ({\bf x}),
\end{equation}

\begin{equation}
j_{op} ({\bf g}) \psi ({\bf x}) = \frac{\hbar}{2im} {{\bf g (x).\nabla} \psi ({\bf x}) + {\bf \nabla} . [{\bf g (x)} \psi ({\bf x})]} +
 D [{\bf \nabla . g(x)}] \psi ({\bf x}).
\end{equation}
Here D is a real number. These operators are seen to be consistent with a Fokker-Planck type of equation,

\begin{equation}
\frac{\partial \rho}{\partial t} = - {\bf \nabla . j} + D \nabla^2 \rho
\label{f_p}.
\end{equation}
A linear Schr\"{o}dinger equation does not satisfy (\ref{f_p}), but the following nonlinear equation does. This may be derived in a manner 
analogous to that of the derivation of (\ref{seqn}) from (\ref{ceqn}),
\begin{equation}
i \hbar \frac{\partial \psi}{\partial t} = - \frac{\hbar^2}{2m} \nabla^2 \psi + V \psi + i D \hbar \nabla^2 \psi + 
i D \hbar \frac{|\nabla \psi|^2}{|\psi|^2} \psi
\label{DG}.
\end{equation}
This is the Doebner-Goldin equation.
Thus, for the given algebra of probability density and current operators of the wave-function, many inequivalent one-particle representations can be found, 
one of which yields the usual linear Schr\"{o}dinger equation, and yet another one of which yields (\ref{DG}). D may be a measure of the effect of a
 measuring apparatus, negligible when the number of degrees of freedom of the system is small. In our considerations of our nonlinear equation in the next Section, we will have a parameter very similar to $D$, which depends on the mass of the particle for which the nonlinear equation is being written.

Goldin has generalized this equation to a broader class of nonlinear equations \cite{goldin}, by introducing greater symmetry between phase and amplitude of the quantum state. As explained in \cite{singhncomm}, the equation we now consider belongs to this broader Goldin class. 

\section{A Physical Model for Quantum Measurement}

We will now present a model which we regard as being more physical than Grigorenko's, using a special case of the Doebner-Goldin system of equations. This special case is of interest also because it results from an attempt, albeit tentative, to develop a mathematical description of quantum mechanics which does not refer to a classical time, using the language of noncommutative geometry. We will describe the measurement process in a pointer basis, in conformity with the assumption that the measurement of an observable is done by a pointer, which upon performing a measurement goes to a specific point correlated with the eigenstate of the observable. The full system can be thought of as a ``combined''  state of pointer wave-function and wave-function of the quantum system, with the prescription that the pointer position gives information about the state of the quantum system. A major advantage of this treatment is that it is independent of which observable of the quantum system is being measured; also, the role of the measuring apparatus is brought out explicitly. 
At the outset we emphasize that the pointer is {\it not} assumed to be a classical object with a definite position and momentum - this classical property is in fact a consequence of the collapse induced by the nonlinearity, as we will elaborate further below. 
The discussion presented here matches the results obtained by the more heuristic treatment in \cite{singh2007}.
After analyzing the importance of nonlinearity for quantum measurement, we will estimate the time-scales for nonlinearity induced collapse. Finally we will suggest laboratory and thought experiments which could test the idea that collapse of the wave-function is caused due to a nonlinearity.  

Our starting point is the following very interesting model equation, which mimics the Hamilton-Jacobi equation of non-relativistic classical mechanics, for a
particle of mass $m$ : 
\begin{equation}
\frac{\partial S}{\partial t} = - \frac{1}{2m} \left(\frac{\partial S} {\partial q}\right)^{2} + i\hbar\theta\left(\frac{m}{m_{Pl}}\right) 
\frac {\partial^{2}S}{\partial q^{2}}.
\label{hjq}
\end{equation}
This equation is assumed to describe the regime `in-between' quantum mechanics and classical mechanics, i.e. the mesoscopic regime.
Here, $S$ is a {\it complex} function of time and of the configuration variable $q$, and $\theta$ is is a monotonic function of the ratio of the mass $m$ to Planck mass, going from one as $m\rightarrow 0$, to zero as $m\rightarrow \infty$. Setting $\theta=0$ and taking $S$ to be real in this limit converts the above equation to the standard classical Hamilton-Jacobi equation for a free particle. Setting $\theta=1$ and substituting $\Psi=e^{iS/\hbar}$ converts (\ref{hjq}) to the standard linear Schr\"{o}dinger equation. On the other hand, when $\theta$ is neither zero nor one, the substitution $\Psi=\exp{iS}/\hbar$ in (\ref{hjq}) leads to the following nonlinear Schr\"{o}dinger equation  
\begin{equation}
i\hbar \frac{\partial\Psi}{\partial t} = - \frac{\hbar^{2}}{2m} \frac{\partial^{2}\Psi}{\partial q^{2}}+ \frac{\hbar^{2}}{2m}(1-\theta)
\left( \frac{\partial^{2}\Psi}{\partial q^{2}} - \left[ \frac{\partial}{\partial q} \ln \Psi\right]^{2}\Psi\right).
\label{nlse}
\end{equation}

This equation is similar to the simplest of the D-G equations, Eqn. (\ref{DG}) but differs from that equation in important ways. Unlike the D-G equation, it does not satisfy the Fokker-Planck equation. It satisfies the continuity equation, provided the probability density and current are defined using an effective wave-function $\Psi_{eff}$ :
\begin{equation}
\rho=|\Psi_{eff}|^2, \quad j=  -i\frac{\hbar\theta}{2m}\left(\Psi^{*}_{eff}\frac{\psi_{eff}}{\partial q} - \Psi_{eff}
\frac{\Psi_{eff}^{*}}{\partial q}\right),
\quad \Psi_{eff} \equiv \Psi^{1/\theta}.
\label{normp}
\end{equation}

The quasi-Hamilton-Jacobi equation (\ref{hjq}) has the following natural generalization to a two particle system, 
$(m_{1},q_{1})$ and $(m_{2},q_{2})$, in analogy with ordinary quantum mechanics. 
\begin{equation}
\frac{\partial S}{\partial t} = - \frac{1}{2m_{1}} \left(\frac{\partial S} {\partial q_{1}}\right)^{2} + 
i\hbar\theta(m_{1}) 
\frac {\partial^{2}S}{\partial q_{1}^{2}}  - \frac{1}{2m_{2}} \left(\frac{\partial S} {\partial q_{2}}\right)^{2} + i\hbar\theta(m_2) 
\frac {\partial^{2}S}{\partial q_{2}^{2}}.
\label{hjq2}
\end{equation}
A two-particle equation such as this one is an appropriate equation to work with, for describing the measurement process. We will assume that $(m_{2},q_{2})$ is the quantum system, and $(m_{1},q_{1})$ is the measuring apparatus. Furthermore, since $m_{2}\ll m_{Pl}$ we will set $\theta_{2}=1$. 
With the substitution $\Psi=\exp(iS/\hbar)$ in (\ref{hjq2}) the two-particle nonlinear Schr\"{o}dinger equation for describing the interaction of the quantum system with the measuring apparatus is
\begin{equation}
i\hbar \frac{\partial\Psi}{\partial t} = - \frac{\hbar^{2}}{2m_{1}} \frac{\partial^{2}\Psi}{\partial q_{1}^{2}} - \frac{\hbar^{2}}{2m_{2}} \frac{\partial^{2}\Psi}{\partial q_{2}^{2}}+ \frac{\hbar^{2}}{2m_{1}}(1-\theta_{1})
\left( \frac{\partial^{2}\Psi}{\partial q_{1}^{2}} - \left[ \frac{\partial}{\partial q_{1}} \ln \Psi\right]^{2}\Psi\right).
\label{nlse2}
\end{equation}   
We will comment below on the question of associating a conserved norm with this equation. Before doing so, we provide a careful interpretation of the measurement process, in the pointer basis. 

We assume that the state of the apparatus is strongly coupled to the state of the quantum system. The quantum system interacts with the apparatus and this leads to a combined state. If the quantum system is in a definite eigenstate $|\phi_n>$ and interacts with the apparatus we get an outcome 
$|A_{n},\phi_n> = |A_{n}>|\phi_n>$, with $|A_{n}>$ being the resultant apparatus state (marked by its pointer). 
Therefore, generically the apparatus should act on the quantum state $ |\Phi_n> = \sum_n a_n|\phi_n> $ to give the state 
$ \sum_n a_{n}|\phi_n>|A_{n}>$.

Now, correlation between the apparatus and the quantum system is such that $a_n$ is related to the wave function $\psi(x)$ of apparatus pointer
 in position space. The position of the pointer is related to the state of the quantum system through the following interpretation :
There are ranges of position of the pointer which depend on the eigenstates of the quantum system. If the pointer is between $x_n$ and $x_n + \Delta x_n$, one concludes that the quantum system is in the $n^{th}$ eigenstate.
Each pointer position is related with an eigenstate of the quantum system as 
$$a_n|A_n> = \int_{\Delta x_n}dx \psi(x) |x>.$$
It leads to 
\begin{equation}
 |a_n|^2=\int_{\Delta x_n} |\psi(x) |^2 dx,
\end{equation}
suggesting that if 
$$|a_n|^2 = \delta_{ni}$$ 
then
$$ |\psi(x) |^2 =\delta(x-x_i). $$
This is consistent with the observed fact that if the quantum system is in one definite eigenstate the apparatus pointer goes to the corresponding position with certainty. Alternatively, we can say that the pointer could be between $ x_{n} $ and $ x_{n}+\Delta x_{n} $ ``if and only 
if'' the quantum system is in  $ n^{th} $ eigenstate.

If $|\psi(x) |^2 $ does not vary significantly in an interval $\Delta x$, as in the case where range of position of the pointer is correlated with one eigenstate, then
\begin{equation}
 |a_n|^2 \approx \Delta x_n |\psi(x_n) |^2 .
\end{equation} 

In the abstract basis the combined wave function, when the apparatus couples with the quantum system, can be written as 
$$ |\Psi>  = \left(\sum_n a_{n}|\phi_n>\right)\left(|A> \right) \rightarrow \sum_n a_{n}|\phi_n>|A_n> , $$
with initially  
\begin{equation}
< \Psi|\Psi> =\sum_n |a_{n}|^{2}=1.
\label{sumeqn}
\end{equation}
Identifying apparatus position basis as $x$ and quantum system position as $y$ and writing $|\phi_n> =\int dy \phi_n(y)|y>$ , we have 
\begin{equation}     
|\Psi>  = (\sum_n a_{n}|\phi_n>|A_n> )= \sum_n  \int \int_{\Delta x_n} dxdy \phi_n(y) \psi(x\approx x_n)|x>|y>.
\label{stateexp}
\end{equation}
(This scheme is applicable to those states only which have a position state representation. For example, `spin states' would have to be dealt with differently.) Using
\begin{equation}
 <A_m|A_n> = \delta_{mn},
\end{equation}
\begin{equation}
 <\phi_m|\phi_n> = \delta_{mn},
\end{equation}
we get,
\begin{equation}
 |<\phi_m,A_m|\Psi>|^2 = |a_{m}|^2 .
\end{equation}

We now demonstrate how the nonlinear Schr\"{o}dinger equation (\ref{nlse2}) can explain the collapse of the wave-function.
Consider the following state
\begin{equation}
 \Psi_n = \psi(x)\phi_n(y)
\label{waveeqn} 
\end{equation}
using which one can evidently write
\begin{equation}
 \Psi(x,y) = \sum_n \int_{\Delta x_n} dx \Psi_n(x,y).
\label{xy}
\end{equation}
This is the entangled state for which one must demonstrate the breakdown of superposition upon measurement. Making the substitution
\begin{equation}
\Psi \equiv \Psi_{F} \equiv R_{F}\exp\frac{i\phi_{F}}{\hbar},
\label{ampphase} 
\end{equation}
in terms of amplitude and phase we can write Eqn. (\ref{nlse2}) as
\begin{equation}
i\hbar \frac{\partial\Psi_{F}}{\partial t}=H_{F}\Psi_{F} + 
\frac{\hbar\gamma(m_1)}{2m_{1}}\left(\hbar\frac{\partial^{2}\ln R_{F}}{\partial x^{2}
}+i \frac{\partial^{2} \phi_{F}}{\partial x^{2}}\right) \Psi_{F},
\label{ev2eqn}
\end{equation}
where $$ H_{F} = -\frac{\hbar^{2}}{2m_{1}}\frac{\partial^{2} }{\partial x^{2}} -\frac{\hbar^{2}}{2m_{2}}\frac{\partial^{2} }
{\partial y^{2}} $$
and $ \gamma = 1-\theta_{1} $.

Substituting equation (\ref{waveeqn}) in (\ref{ev2eqn}) to obtain contribution from a single component in (\ref{xy}) gives
\begin{eqnarray}
 i\hbar \frac{\partial\psi(x)}{\partial t}\phi_n(y) & + & i\hbar \frac{\partial\phi_n(y)}{\partial t}\psi(x)= \nonumber  \\
H_{F}\psi(x)\phi_n(y) & + & \frac{\gamma(m_{1})}{2m_{1}}\hbar^{2}\left(\frac{\partial^{2} \ln R_{F}}
{\partial x^{2}}+i\frac{1}{\hbar}\frac{\partial^{2}\phi_{F}}{\partial x^{2}}\right)\psi(x)\phi_n(y).
\label{realeqn}
\end{eqnarray}
Taking complex conjugate of Eqn. (\ref{realeqn}) gives
\begin{eqnarray}
-i\hbar \frac{\partial\psi^{*}(x)}{\partial t}\phi_n^{*}(y) & - & i\hbar \frac{\partial\phi_n^{*}(y)}{\partial t}\psi^{*}(x)= \nonumber \\
H_{F}\psi^{*}(x)\phi_n^{*}(y) & + & \frac{\gamma(m_{1})}{2m_{1}}\hbar^{2}\left(\frac{\partial^{2} \ln R_{F}}{\partial x^{2}}-
i\frac{1}{\hbar}\frac{\partial^{2} \phi_{F}}{\partial x^{2}}\right)\psi^{*}(x)\phi_n^{*}(y).
\label{imagneqn}
\end{eqnarray}
Multiplying (\ref{realeqn}) with $\psi^{*}(x)\phi_n^{*}(y)$ and integrating the variable $ y $ over all space, while $ x $ over a 
small region $ \Delta x $ gives
$$ i\hbar \int_{\Delta x} \psi^{*}(x) \frac{\partial \psi(x)}{\partial t}dx\int_{-\infty} ^{\infty}|\phi_n(y)|^{2}dy + $$
$$ i\hbar \int_{-\infty} ^{\infty} \phi_n^{*}(y) \frac{\partial \phi_n(y)}{\partial t}dy\int_{\Delta x}|\psi(x)|^{2} dx = $$ 
$$ -\frac{\hbar^{2}}{2m_{1}}  \int_{\Delta x}\psi^{*}(x) \frac{\partial^{2} \psi(x)}{\partial x^{2}}dx\int_{-\infty}^{\infty}|\phi_n(y)|^{2} dy $$
$$-\frac{\hbar^{2}}{2m_{2}} \int_{-\infty}^{\infty}\phi_n^{*}(y)\frac{\partial^{2} \phi_n(y)}{\partial y^{2}}dy\int_{\Delta x}|\psi(x)|^{2}dx +$$
\begin{equation}
\frac{\gamma(m_{1})}{2m_{1}} \hbar^{2}  \int_{\Delta x} \psi^{*}(x) \left(\frac{\partial^{2} \ln R_{F}}{\partial x^{2}} +
i\frac{1}{\hbar}\frac{\partial^{2} \phi_{F}}{\partial x^{2}}\right) \psi(x)dx \int_{-\infty} ^{\infty}  |\phi_n(y)|^{2} dy.
\label{ev3eqn}
\end{equation}

Similarly, multiplying (\ref{imagneqn}) with $\psi(x)\phi_n(y)$ and integrating the variable $ y $ over all space, while $ x $ over a 
small region $ \Delta x $ gives

$$ - i\hbar \int_{\Delta x}\psi(x) \frac{\partial \psi^{*}(x)}{\partial t}dx \int_{-\infty} ^{\infty} |\phi_n(y)|^{2}dy $$
$$ - i\hbar \int_{-\infty} ^{\infty} \phi_n(y)\frac{\partial \phi_n^{*}(y)}{\partial t}dy\int_{\Delta x} |\psi(x)|^{2} dx = $$
$$ - \frac{\hbar^{2}}{2m_{1}}\int_{\Delta x} \psi(x)\frac{\partial^{2} \psi^{*}(x)}{\partial x^{2}}dx \int_{-\infty}^{\infty} |\phi_n(y)|^{2} dy $$
$$ -\frac{\hbar^{2}}{2m_{2}}\int_{-\infty}^{\infty} \phi_n(y)  \frac{\partial^{2} \phi_n^{*}(y)}{\partial y^{2}}dy\int_{\Delta x} |\psi(x)|^{2} dx +$$
\begin{equation}
\frac{\gamma(m_{1})}{2m_{1}}\hbar^{2}\int_{\Delta x} \psi(x) \left(\frac{\partial^{2} \ln R_{F}}{\partial x^{2}}-i \frac{1}{\hbar}
\frac{\partial^{2}\phi_{F}}{\partial x^{2}}\right)\psi^{*}(x) dx\int_{-\infty} ^{\infty} |\phi_n(y)|^{2} dy.
\label{ev4eqn}
\end{equation}

Noting that the quantities 
$$ \psi(x),\psi''(x),\frac{\partial^{2} \ln R_{F}}{\partial x^{2}},\frac{\partial^{2} \phi_{F}}{\partial x^{2}} $$ 
do not vary appreciably within a small 
interval $ \Delta x $, and subtracting (\ref{ev4eqn}) from (\ref{ev3eqn}) we get
\begin{eqnarray}
i\hbar \frac{\partial}{\partial t}\left(\int_{\Delta x} \psi(x)\psi^{*}(x)dx\int_{-\infty} ^{\infty}\phi_n^{*}(y)\phi_n(y)dy\right) & = &\nonumber\\
- \frac{\hbar^{2}}{2m_{1}}\int_{\Delta x} { \left(\psi^{*}(x)\frac{\partial^{2} \psi(x)}{\partial  x^{2}}-\psi(x)\frac{\partial^{2}
 \psi^{*}(x)}{\partial x^{2}}\right) } dx \int_{-\infty}^{\infty} |\phi_n(y)|^{2} & dy & \nonumber\\
+ 2i\frac{\gamma(m_{1})}{2m_{1}}\hbar\int_{\Delta x}
\psi(x)\frac{\partial^{2} \phi_{F}}{\partial x^{2}}\psi^{*}(x)dx\int_{-\infty} ^{\infty}|\phi_n(y)|^{2} & dy & ,
\label{teveqn}
\end{eqnarray}
or, equivalently,
$$ i\hbar\frac{\partial}{\partial t}\left(\psi(x)\psi^{*}(x)\Delta x\int_{-\infty} ^{\infty}\phi_n^{*}(y)\phi_n(y)dy\right)=$$
$$\left[-\frac{\hbar^{2}}{2m_{1}}\frac{ \psi^{*}(x)\psi'' (x)-\psi^{*}{''} (x)}{|\psi(x)|^{2}}+2i\frac{\gamma(m_{1})}{2m_{1}}
\hbar \frac{\partial^{2} \phi_{F}}{\partial x^{2}}\right]$$
\begin{equation}
|\psi(x)|^{2}\Delta x\int_{-\infty} ^{\infty}|\phi_n(y)|^{2}dy.
\end{equation}

Now, since $ |\psi(x)|^{2}\Delta x\int_{-\infty} ^{\infty}|\phi_n(y)|^{2}dy $ is  equal to $|a_{n}|^{2} $ for $ x = x_{n} $ and for normalized basis states $ \phi_n(y)$, the equation (\ref{teveqn}) can be re-written as
\begin{equation}
 \frac{\partial |a_{n}|^{2} }{\partial t} = Q(x_{n})|a_{n}|^{2},
\label{popeqn}
\end{equation}
where 
\begin{equation}
  Q(x) = \frac{\hbar}{m_{1}}\left[\gamma \frac{1}{\hbar} \frac{\partial^{2} \phi_{F}}{\partial x^{2}}-
\frac{Im [\psi^{*}(x)\psi''(x)]}{|\psi(x)|^{2}}\right].
\end{equation}

Thus, starting from the nonlinear Schr\"{o}dinger equation (\ref{nlse2}) we have arrived at the crucial equation (\ref{popeqn}) which determines the effect of the nonlinearity on the initial quantum superposition present in the quantum system $(q_{2},m_{2})$ at the onset of measurement. 
To make further progress, we will make two important, physically motivated, approximations. Firstly, as we have reasoned in Section 2, when the mass of the system in question [here quantum system plus apparatus] becomes comparable to Planck mass, the nonlinearity dominates the evolution. Hence in Eqn. (\ref{popeqn}) above, we will drop the term resulting from having retained the linear part in (\ref{nlse2}). It is easily shown that (\ref{popeqn}) retains its form, but the second term in $Q(x)$ can be dropped, and $Q(x)$ simplifies to
\begin{equation}
  Q(x) = \frac{\gamma }{m_{1}}\frac{\partial^{2} \phi_{F}}{\partial x^{2}}.
\label{qsimp}
\end{equation}
Secondly, phases are expected to vary significantly over a length scale much smaller than that of amplitude variation, with the condition 
 $$\hbar \frac{\partial^{2} \phi_{F}}{\partial x^{2}} \gg \hbar^2\frac{\partial^{2} \ln R_{F}}{\partial x^{2}}.$$
This premise is self-consistent with the results, because in the end we conclude that the `apparatus plus quantum system' together
become localized, a property true of classical systems, for which it is in turn known, from the semiclassical [WKB] limit, that the
phase varies much more rapidly, compared with the amplitude.
As a result the Hamiltonian is completely imaginary and it can be shown that phase becomes constant in time in such an evolution
 \begin{equation}
i\hbar \frac{\partial\Psi_{F}}{\partial t}=i\hbar \frac{\gamma(m_{1})}{2m_{1}}\frac{\partial^{2} \phi_{F}}{\partial x^{2}}\Psi_{F}.
\end{equation}
Writing $$ \Psi \equiv \Psi_{F} =R_{F}\exp\frac{i\phi_{F}}{\hbar}, $$ one can see that $$ \frac{\partial\phi_{F}}{\partial t} = 0 .$$ 
Therefore, at the moment the classical apparatus (hence nonlinear term) takes over the linear part, the phase of the wave function becomes 
practically fixed (in time). Hence $Q$ becomes constant, taking the value at the onset of measurement.
We also note that when the nonlinear part is dominant, the nonlinear Schr\"{o}dinger equation (\ref{nlse2}) essentially behaves like the single particle nonlinear equation, which {\it is} norm preserving, in accordance with (\ref{normp}). Thus the sum $\sum |a_{n}|^2$ will be preserved during the nonlinear evolution. It is not clear though, how to associate a conserved norm with Eqn. (\ref{nlse2}).

The nonlinear evolution is now exactly as in Grigorenko's model. If we consider two quantum states $|\phi_i>$ and $\phi_j>$, their population ratio evolves as
\begin{equation}
 \frac{\partial }{\partial t}\ln \left(\frac{|a_{i}|^{2}}{|a_{j}|^{2}}\right) = Q(x_{i})-Q(x_{j}).
\label{ratio}
\end{equation}
Thus the system will evolve to that state for which the value of $Q$ is the largest at the onset of measurement, thereby breaking superposition.
After the evolution is complete the wave function is such that only one of the
$$ |a_{i}|^{2} \equiv \left( \int_{-\infty} ^{\infty}|\phi_n(y)|^{2}dy \right) |\psi(x)|_{x=x_{i}} ^{2}\Delta x_{i} $$ 
say $i= n$, becomes very strongly dominant in contribution to (\ref{sumeqn}), as seen from (\ref{ratio}). That corresponds to saying that the pointer tends to be present between $x_{n}$
and $ x_{n} +\Delta x_{n}$ 
which itself corresponds to the  $ n^{th} $ eigenstate of the quantum system. 

The question that arises next is the derivation of the Born probability rule, for which it is essential that the $Q$s be random variables. For us,
the $Q$s are guaranteed to be random variables, because according to their definition (\ref{qsimp}) they are related to the random phase $\phi_F$
of the overall composite quantum state of the apparatus and system, as introduced in Eqns. (\ref{stateexp}) and (\ref{ampphase}). Of course, the $Q$'s are not directly related to the phase itself, but the value of its second gradient, at the location of the pointer. 

Now, following Grigorenko, if the probability distribution of $Q$ at the instant of measurement is given as 
$$ \omega (Q_n) = |<\Psi (t_0)|\phi_n>|^2 e^{|<\Psi (t_0)|\phi_n>|^2 Q_n} $$  in the interval $(-\infty,0 ]$ for the random variables $Q_n$, we get 
the probability of finding $ Q_i $ as the largest among all $Q'$s as
$$ p_i = \int .. \int \omega(Q_i) dQ_i \prod_{n \neq i} \theta(Q_i - Q_n) \omega(Q_n) dQ_n
\label{p_i} = |<\Psi(t_0)|\phi_i>|^2 .  $$
This relation could well be expressed in position space and $ |\phi_n> $ here 
are the eigenstates (of the operator) the quantum system could be in.
Thus, we can recover the Born rule. To eliminate the dependence of the probability distribution of the random variable on the initial state
 we make the transformation to a new random variable $u_n$ as before,
$$ Q_n =\frac{u_n}{|<\Psi|\phi_n>|^2}, $$
with the new random variable $ u_n $ distributed along $(-\infty,0  ]$ as
$$\omega (u_n) = e^{u_n}. $$
The evolution equation is then written as,
\begin{equation}
 \frac{\partial \ln(|a_{n}|^{2}) }{\partial t} = Q_n,
\end{equation}
or,
\begin{equation}
 \frac{\partial \ln(|a_{n}|^{2}) }{\partial t} = \frac {u_n}{|a_{n}|^{2}}.
\end{equation}

Now, the question to be asked is what is the probability distribution of $Q_n$ at the time of measurement, given its dependence on  random
 variables $u_n$ and the probability distribution of $u_n$?
We obtain that the random variables $u_n$ give rise to the required probability distribution for $Q_n$ at $ t=t_0 $, i.e. at the time of measurement. As in Grigorenko's case, the most natural choice for $u_n$ would be that they are related to the phases $\chi_n$ of the eigenstates 
$|\phi_n>$ of the quantum system, via $u_n=\ln(\chi_n/2\pi)$ with $\chi_n$ uniformly distributed in $[0, 2\pi]$. For us, using the definition of $Q_n$, this gives the relation
\begin{equation}
1-\theta(m_1) = \frac{m_{1}\ln(\chi_{n}/2\pi)}{|a_n|^2 \phi_{F}''(x=x_n)}
\end{equation}
As of now, this relation must be treated as an {\it ad hoc} aspect of our model. What it says is that if the Born probability rule is to be a consequence of the random distribution of phases, this relation must hold. Now, in our model, we do not have an underlying theory for the relation of the function $\theta(m_1)$ to quantum states. Only after one has such a theory, can the validity of this relation be verified. Until then, the exact derivation of the Born rule in our model has to be considered as being tentative, although evidence for randomness is already there in the model.

It should be mentioned that the simplest D-G equation, Eqn. (\ref{DG}), does not model wave-function collapse. It turns out that in this case, the dominant term of the Hamiltonian at the onset of measurement is Hermitian, instead of being non-Hermitian, and hence cannot bring about exponential growth or decay of stationary states. We also note that it is possible to construct a relativistic version of our nonlinear Schr\"{o}dinger equation
\cite{singhncomm}. 

\noindent{\bf Nonlinearity and Superluminality} : There has been considerable discussion on the issue of superluminality in the case where the evolution of a quantum system, before and during measurement, is described by a deterministic equation. Firstly, it was shown by Gisin \cite{gnl}, \cite{gnl2} that such a deterministic evolution must be nonlinear. Further, assuming that the linear projection postulate continues to hold in the nonlinear theory, Gisin showed that superluminal propagation is possible. However, various authors have shown that this assumption need not always hold \cite{polchinski}, \cite{lucke}, and hence that there are nonlinear equations which do not admit superluminal propagation. 
Czachor and Doebner \cite{czachor} showed that faster-than-light communication could be avoided by appropriately defining a nonlinear version of the projection postulate.
Luecke also showed \cite{lucke} that a general class of Doebner-Goldin equations is consistent with the requirement of causality. Since our model belongs to the general D-G class, this issue should be regarded as open for now, and addressing it is non-trivial, for the following reasons.

One will have to develop a position independent description in the form of density matrix dynamics, to see the form of consistent projection operators, and hence obtain the form of observables from it. Grigorenko has argued that it might not always be possible to develop closed form density matrix formalism for nonlinear evolution and hence one cannot be sure of the existence of a superluminal world-line. Also, when nonlinearity sets in at a particular instant of time and is not present throughout the evolution, it is not clear how to construct a superluminal world-line. It would also be relevant to recall that in our case the nonlinearity, and hence the related issue of superluminality (or otherwise) both arise in a domain that is currently untested by experiment.  

We would also like to mention that perhaps our proposed reformulation of quantum mechanics, in a manner so that no reference is made to an external classical time, can help understand the otherwise peculiar presence of EPR correlations outside the light-cone.  In the reformulation, since there is no classical time, there is no light cone. It is then no surprise that such correlations can exist. Only when seen from the vantage point of an external classical spacetime, where there is a light cone, do the correlations appear peculiar.

\section{The lifetime of a quantum superposition : towards experimental tests}

Eqn. (\ref{popeqn}) defines a time-scale for collapse of the wave-function, during a quantum measurement, which is given by
\begin{equation}
\tau=\frac{1}{Q}=\frac{m_1}{[1-\theta(m_1)]\phi_F''}.
\label{lifetime}
\end{equation} 
If our model is to be trusted as a realistic description of measurement, this expression should yield realistic numbers. We can get a rough estimate of the magnitude of $\tau$ by noting that since $m_1$ is comparable to or much larger than Planck mass, $\theta$ can be taken to be close to zero. An order of magnitude estimate for $\phi''$ is $\phi/L^{2}$, where $L$ is the extent of the linear dimension of the measuring apparatus. For estimating the magnitude of the phase we can take the phase to be equal to the classical action (leading order contribution) $m_{1}c^{2}t$, where $t$ is the time for which we observe the classical trajectory of the measuring apparatus. Assuming $t$ to be of the order of $\tau$ itself, and putting things together, we get that
\begin{equation}
\tau\approx\frac{L}{c}.
\label{lbyc}
\end{equation}    
If the measuring apparatus is assumed to have a linear dimension of say $1$ mm, this time is around $10^{-11}$ seconds, which seems small enough to be consistent with experiments.  

This expression for the time-scale can be used to make an important prediction, which departs from standard quantum mechanics, and which may become testable through ongoing and future experiments. Our model says that when the mass of a system approaches Planck mass, the nonlinearity in the nonlinear Schr\"{o}dinger equation becomes important, and causes an initial linear quantum superposition to break down, over a finite time scale. This would explain then, why macroscopic objects are never found in superposition of position states - the superposition lifetime is extremely small. On the other hand, when one considers mesoscopic objects, the superposition lifetime would be large, but perhaps detectable. Only when one starts considering atomic systems, the superposition lifetime becomes astronomically large, and the linear superposition principle is experimentally observed to hold to a very high accuracy. As seen from Eqn. (\ref{lifetime}) the lifetime goes to infinity as $m_1$ goes to zero, since in that limit $\theta(m_1)$ goes to one. In other words as the mass [equivalently number of degrees of freedom] of the object is increased from atomic values to macroscopic values, the superposition lifetime comes down from astronomically large (and unmeasurable) values to extremely small (and again unmeasurable) values. Somewhere in between then, the lifetime should become measurable - this is expected to happen in the mesoscopic domain. The collapse of the wave-function of an object into one particular position state is again described by the nonlinear Eqn. (\ref{popeqn}) and the associated superposition lifetime is given by (\ref{lifetime}). The physical process that explains a quantum measurement is the same as the one that explains absence of superposition in macroscopic objects.

To get an estimate of $\tau$ in the mesoscopic domain, we will use an expression for the mass dependence of $\theta(m_1)$ which we have argued for elsewhere \cite{singhdual}
\begin{equation}
\theta(m_1) = \frac{1}{1+\frac{m_{1}}{m_{Pl}}}.     
\end{equation}
We approximate $\phi''$ as $\phi/L^{2}$ as before. For the magnitude of the phase, we assume a conservative ground state estimate $\phi=N\hbar$, where $N$ is the number of atoms in the mesoscopic object. A simplistic estimate for $L$ would be to take $L^{3}\sim N a^{3}$, where $a$ is the linear dimension of an atom. Then we can rewrite the expression (\ref{lifetime}) for the superposition lifetime as
\begin{equation}
\tau \approx \left(1+\frac{m_1}{m_{Pl}}\right) \frac {m_{Pl} \ a^{2}} {\hbar} \frac {1} {N^{1/3}}
\end{equation}
The ratio $m/m_{Pl}$ in the first bracket is ignorable for mesoscopic objects, for which we take $N \ll 10^{18}$ (and hence $m \ll m_{Pl}$).
Taking $a$ to be one Angstrom, we get $\tau \approx 10^{6}/N^{1/3}$ seconds. Thus, for $N= 10^{9},\ 10^{12},\ 10^{15}$, the superposition lifetime is $10^{3}$, $10^{2}$, $10$,  seconds respectively. It is also easily verified that for $m\gg m_{Pl}$, where one can approximate $\theta$ to zero,  the superposition lifetime is again given by $L/c$ and is hence extremely small.

Strictly speaking, the result of Eqn. (\ref{lbyc}) is expected to hold only in the vicinity of Planck mass - that is the domain where the specific nonlinear approximation made by us holds. We can get a somewhat better prediction, using the form of $\theta(m_1)$ suggested above, and taking the phase as $mc^{2}\tau$ we can see that
\begin{equation}
\tau^{2} \approx \frac{m_{Pl}a^{2}}{m_{atom}c^{2}}  \frac{1}{N^{1/3}}
\end{equation}
which for $N=10^{18}$ gives $\tau = 10^{-12}$ seconds, consistent with expectation.

While our nonlinear Schr\"{o}dinger equation is only a model, and while our numerical estimates are rather rough, the results suggest that linear superposition will be put to experimental test in the near future, by study of objects made of about $10^{12}$ atoms. In this context, we would like to mention three very impressive ongoing experiments which are working towards building quantum superposition of large [mesoscopic] objects by minimizing the effects of decoherence by cooling the objects to extremely low temperatures. Our prediction is that [assuming decoherence effects have been adequately eliminated] it will indeed be possible to build a  quantum superposition, but it will last for only a finite time, in accordance with the lifetimes given in the previous paragraph. Thus in order to test the validity of the linear superposition principle for a mesoscopic object, and to rule out the nonlinear theory, it is not enough to make the superposition; it has to be shown to last for a sufficiently long time. Otherwise, one will only be putting a bound on the validity of the principle.  The nonlinear theory will be verified for an object if it can be shown that collapse occurs on a time-scale smaller the decoherence time-scale.
 
\smallskip
 
\noindent {\bf The Vienna Experiment} : The idea is to strongly couple photons to a micro-mechanical resonator [having say a billion atoms] so that
sufficient energy can be transferred between the photons and the resonator. The photons are themselves in a superposition of two different  
states - this will result [because of entanglement] in the resonator being in a superposition of a state `here' and a state `there' \cite{vienna},
\cite{vienna2}.

\smallskip

\noindent{\bf The CalTech Experiment} : The experiment aims to create superposition of position states of nanomechanical oscillators by coupling them to superconducting qubits \cite{caltech}. 

\smallskip

\noindent{\bf The LIGO Science Collaboration Experiment} : An interferometric gravitational wave detector - a kilogram scale object - has been cooled to an effective temperature of a micro-Kelvin at the Laser Interferometer Gravitational-wave Observatory at MIT. If the linear extent of the object is 10 cms we
predict that the superposition lifetime will be about $10^{-8}$ seconds. Of course, the system has to be cooled sufficiently so that the deoherence lifetime is more than $10^{-8}$ seconds \cite{mit} 

\smallskip

\noindent{\bf Proposed Oxford Experiment} : It has been proposed that quantum superposition states of an object involving about $10^{14}$ atoms can be created using existing technology, through the interaction of a photon with a mirror. The mirror is part of an optical cavity which forms one arm of an interferometer. By observing the photon, one can infer the creation of superposition and decoherence states involving the mirror \cite{oxford}.

\smallskip

\noindent{\bf A Thought Experiment} : If the input quantum state for the apparatus is
$$ |\psi_{in} > =\sum_n a_n |\phi_n >, $$
and $$ <\psi_{in}|\psi_{in} > = \sum_n |a_n|^2 <\phi_n |\phi_n >  $$
the nonlinear Schr\"{o}dinger equation predicts that the outcome from the apparatus after a time $t$ will be  
\begin{equation}
  < \psi_{out}|\psi_{out} >_t =  \left(1 + \sum_n \exp(-\gamma Q_{mn}t) \right)|a_m(t)|^2 <\phi_n |\phi_n >,
\label{partial}
\end{equation}
where
$$ |a_m(t)|^2  = |a_m(t=0)|^2 \exp(\gamma Q_{m}t), $$
and $ Q_{mn} = Q_m-Q_n $, and we have assumed that $ Q_m $ was the largest amongst the $Q$'s, at $ t=0$.

On the other hand, according to standard quantum mechanics the outcome is  
\begin{equation}
  < \psi_{out}|\psi_{out} > =  \sum_n (\delta_{nm} <\phi_n |\phi_n >).
\end{equation}
Standard quantum mechanics predicts that the outcome from any apparatus is only one eigenstate
$$ |\psi_{out} > = \sum_n \delta_{nm}|\phi_n > ,$$
while in this nonlinear case superposition structure is retained in the outcome for a period small compared to the collapse lifetime $\tau$,
$$|\psi_{out} >_t = \left(|\phi_m > + \sum_n \exp(-\gamma Q_{mn}t/2)|\phi_n > \right)|a_m(t)|  ,$$
apart from phases in the coefficients.

Suppose then that one makes a `partial' first measurement, where one introduces the measuring apparatus in the path of the quantum system for a time-scale that is much smaller than $\tau$. The outcome will be the state given by Eqn. (\ref{partial}), a state which is of course not observed
in ordinary quantum mechanics. We can infer the existence of such a state by feeding it into a second measuring apparatus and this time performing a complete measurement, i.e. keeping the second apparatus in place for a time much longer than $\tau$. The outcome of the second measurement will evidently be different from what quantum mechanics predicts for a pair of successive measurements.

\section{Outlook}

We have presented the case here that nonlinearity in quantum mechanics is inevitable in the mesoscopic domain, and that the linear theory is an excellent approximation in the microscopic limit. In future work we will try to arrive at more precise estimates for the lifetime of a quantum superposition, and also investigate the practical feasibility of sufficiently eliminating decoherence so that the possible presence of a nonlinearity can be verified or ruled out. It would also be interesting to compare these ideas with the phenomenological model of Continuous Spontaneous Localization \cite{grw}, \cite{angelo} and see if ideas similar to those presented here can provide a theoretical underpinning  
for the CSL model.

\bigskip

\noindent{\bf Acknowledgement} : We thank Aruna Kesavan for useful discussions during the early stages of this work. The discussion and presentation in Sections 3 and 4 is based on her Masters Thesis [December, 2008].


\begin{thebibliography}{99}
\bibitem{zeillinger0} Arndt M et al. 1999 {\it Nature} {\bf 401} 680
\bibitem{zeillinger} Hackerm\"{u}ller et al. 2003 {\it Phys. Rev. Lett.} {\bf 91} 090408
\bibitem{bohmbub} Bohm D and Bub J 1966 {\it Rev. Mod. Phys.} {\bf 38} 453
\bibitem{pearle} Pearle P 1976 {\it Physical Review D} {\bf 13} 857
\bibitem{feynman} Feynman R P 1957 {\it Lectures on Gravitation}
\bibitem{grigorenko} Grigorenko A N 1995 {\it J. Phys. A: Math. Gen.} {\bf 28} 1459
\bibitem{adler} Adler S L 2004 {\it Quantum Theory as an Emergent Phenomenon} Cambridge University Press  
\bibitem{grw} Ghirardi G C, Rimini A and Weber T {\it Physical Review D} {\bf 34} 470
\bibitem{diosi} Diosi L 1989 {\it Physical Review A} {\bf 40} 1165 
\bibitem{penrose} Penrose R 1996 {\it Gen. Rel. Grav.} {\bf 28} 581
\bibitem{singh2007} Singh T P 2009 {\it J. Phys. Conf. Series} {\bf 174} 012024
\bibitem{weinberg} Weinberg S 1989 {\it Annals of Physics} {\bf 194} 336
\bibitem{Einstein} Einstein A 1914 {\it Koniglich Preussische Akademie der Wissenschaften (Berlin) Sitzungberichte} 1030  
\bibitem{christian} Christian J 1998 in {\it Physics meets philosophy at the Planck scale} [arXiv:gr-qc/0108040]
\bibitem{singhncomm} Singh T P 2006 {\it Bulg. J. Phys.} {\bf 33} 217 [arXiv: gr-qc/0510042]
\bibitem{Gisin} Gisin N 1981 {\it J. Phys. A: Math. Gen.} {\bf 14} 2259 
\bibitem{DG} Doebner H D and Goldin G A 1982 {\it Phys. Lett. A} {\bf 162} 397
\bibitem{Sharp} Dashen R F and Sharp H D 1968 {\it Physical Review} {\bf 165} 1857
\bibitem{goldinref} Goldin A 1971 {\it J. Math. Phys.} {\bf 12} 462; Goldin G A and Sharp H D 1970 in {\it Group Representations in Mathematics and Physics : Battele-Seattle Recontres 1969, Lecture Notes in Physics 6, Berlin : Springer-Verlag} 300-310  
\bibitem{goldin} Goldin G A 2000 {\it quant-ph/0002013}; H.-D. Doebner and G A Goldin, 1996 {\it Phys. Rev.} {\bf A54} 3764. 
\bibitem{singhdual} Singh T P 2008 {\it Gen. Rel. Grav.} {\bf 40} 2037
\bibitem{gnl} Gisin N 1989 {\it Helvetica Physica Acta} {\bf 62} 363
\bibitem{gnl2} Gisin N and Rigo M 1995 {\it J. Phys. A. : Math. Gen.} {\bf 28} 7375
\bibitem{polchinski} Polchinski J 1991 {\it Phys. Rev. Lett.} {\bf 66} 397
\bibitem{czachor} Czachor M and Doebner H.-D 2002 {\it Phys. Lett. A} {\bf 301} 139 [arXiv:quant-ph/0110008]
\bibitem{lucke} Lucke W 1999 arXiv:quant-ph/990401
\bibitem{vienna} Gr\"{o}blacher S, Hammerer K, Vanner M R and Aspelmeyer M 2009 {\it Nature} {\bf 460} 724
\bibitem{vienna2} Gr\"{o}blacher et al. 2009 {\it Nature Physics} {\bf 5} 485
\bibitem{caltech} Schwab K et al. 2009 {\it http://www.kschwabresearch.com/articles/detail/8}
\bibitem{mit} Abott B et al. 2009 [LIGO Science Collaboration] {\it New J. Phys.} {\bf 11} 073032
\bibitem{oxford} Marshall W et al. 2003 {\it Phys. Rev. Lett.} {\bf 91} 130401 
\bibitem{angelo} Adler S L and Bassi A 2009 {\it Science} {\bf 325} 275 [arXiv:0912:2211 [quant-ph]]
\end{thebibliography}
\end{document}